\documentclass[journal]{IEEEtran}

\renewcommand{\P}{\mathbb{P}}

\newcommand{\beq}{\begin{equation}}
\newcommand{\eeq}{\end{equation}}
\newcommand{\beqa}{\begin{eqnarray}}
\newcommand{\eeqa}{\end{eqnarray}}

\newcommand{\E}{\mathbb{E}}

\newcommand{\N}{{\cal N}}

\usepackage[crop=pdfcrop,cleanup={.tex, .dvi, .ps, .pdf, .log, .aux, .bbl, .out}]{pstool}

\usepackage{cite}

\usepackage{balance}
\usepackage{amsmath, amssymb}
\usepackage{hyperref}

\usepackage{mathtools}
\usepackage{bbm}
\usepackage{algorithm}
\usepackage{algpseudocode}
\usepackage{amsthm}

\newcommand{\RN}[1]{  \textup{\uppercase\expandafter{\romannumeral#1}}}

\usepackage{color}

\usepackage{microtype}

\DeclareMathOperator{\sign}{sign}
\definecolor{bgrd}{rgb}{1,1,1}
\definecolor{grey}{rgb}{0.9,0.9,0.6}
\definecolor{gray}{rgb}{0.5,0.5,0.5}
\definecolor{dkr}{rgb}{0.6,0.2,0.2}
\definecolor{dkg}{rgb}{0,0.5,0}
\definecolor{dkb}{rgb}{0.0,0.1,0.7}
\definecolor{light-gray}{gray}{0.85}

\title{Quickest Detection of COVID-19 Pandemic Onset}

\author{P.~Braca, D.~Gaglione, S.~Marano, L.~M.~Millefiori, P.~Willett and K.~Pattipati

\thanks{P.~Braca, D.~Gaglione and L.~M.~Millefiori  are with the NATO STO CMRE, Research Department, La Spezia, 19126, Italy. S.~Marano is with DIEM, University of Salerno, Italy. P.~Willett and K.~Pattipati are with the ECE Dept., University of Connecticut, Storrs, 06269, USA.
E-mails:
{\small \{paolo.braca, domenico.gaglione, leonardo.millefiori\}@cmre.nato.int,} {\small      marano@unisa.it,} {\small \{peter.willett, krishna.pattipati\}@uconn.edu}}
\thanks{
Peter Willett was supported by AFOSR under contract FA9500-18-1-0463.}
\thanks{The work of Krishna R. Pattipati was supported in part by the U.S. Office of Naval Research, in part by the U.S. Naval Research Laboratory under Grant N00014-18-1-1238 and Grant N00173-16-1-G905, and in part by the Space Technology Research Institutes from National Aeronautics and Space Administration's (NASA's) Space Technology Research Grants Program under Grant 80NSSC19K1076.}}

\begin{document}
\maketitle

\begin{abstract}
This paper develops an easily-implementable version of Page's CUSUM  quickest-detection test, designed to work in certain composite hypothesis
scenarios with time-varying data statistics. 
The decision statistic can be cast in a recursive form and is particularly suited for on-line analysis. 
By back-testing our approach on publicly-available COVID-19 data we find reliable early warning of infection flare-ups, in fact sufficiently early that the tool may be of use to decision-makers on the timing of restrictive measures that may in the future need to be taken.
\end{abstract}


\section{Introduction}

We develop a version of Page's CUSUM quickest-detection procedure~\cite{page,basseville-book,poor-book-quickest,TRUONG2020107299}, applicable to a family of composite-hypothesis changes. We refer to it as MAST --- the mean-agnostic sequential test.
Consider a set of independent Gaussian observations $\{x_n\}$ of constant known standard deviation $\sigma$ and unknown mean sequence $\{\mu_n\}$. At an unknown time, the mean switches from being less than some prescribed limit (but otherwise unknown) to larger than some prescribed limit (but otherwise unknown). The goal is to detect the change, if any, as soon as possible. This framework represents a convenient abstraction of many problems of practical interest. Here we discuss its application to the detection of COVID-19 pandemic waves.

The outbreak of the COVID-19 infection is certainly one of the most serious global crises of the last two decades. The response of the research community was also extraordinary, and comprehensive reviews are recently appearing in the literature~\cite{roberts2021common,hu2020access}. To contain the ``first wave'' of the COVID-19 pandemic in the spring of 2020, strict lockdown measures were imposed in many countries, with huge societal and economic costs~\cite{Anderson2020,Hellewell2020,Maria2020,SHARIF2020,GlobalSupplyChain,millefiori2020covid19}. 
In the fall of 2020, a ``second pandemic wave'' seems to have grown in many regions of the world, and governments and authorities were again faced with the dilemma of if and when to impose social restrictions.
In this work, after developing the MAST quickest detection procedure, we show how it can provide valuable support to make informed and rational decisions, with a focus on detecting the second and subsequent waves of the COVID-19 pandemic.

\section{MAST: A Novel Quickest Detection Test}
\label{sec:MAST}

Along the same lines of the derivations of Page's test, see e.g.,~\cite[Sec. 2.2.3]{basseville-book} or~\cite[Sec. 8.2]{tartakovsky-book}, we consider the following decision problem involving two statistical hypotheses with independent data:
\begin{subequations}
\begin{align}
&\textnormal{null}:  \hspace*{36pt} x_k \sim \N(\mu_{0,k}, \sigma), \hspace*{17pt} k=1,\dots,n, \\ &\textnormal{alternative}: \quad \begin{cases}
x_k \sim \N(\mu_{0,k}, \sigma), & k=1,\dots,j-1,  \\
x_k \sim \N(\mu_{1,k}, \sigma), & k=j,\dots, n.
\end{cases}
\end{align}
\label{eqapp:pageform}%
\end{subequations}
In~\eqref{eqapp:pageform}, $\{x_k\}_{k=1}^n$ are the data available to the decision maker, $1 \le j \le n+1$ is an unknown deterministic change time and the standard deviation $\sigma$ is assumed known. 
Note in \eqref{eqapp:pageform} that in the case $j=n+1$, the alternative hypothesis is equivalent to the null one, i.e., there is no change in regime.
Different from the classical assumption of Page's test, in our problem the expected values before and after the change are unavailable. Accordingly, we model $\{\mu_{0,k}\}_{k=1}^n$ and $\{\mu_{1,k}\}_{k=1}^n$ as unknown deterministic sequences and we assume that they satisfy the following constraints:
\begin{align}
    &\mu_{0,k} \le \delta_{\ell}, \quad  
    \mu_{1,k} > \delta_{u}, \quad 
    0<\delta_\ell \le \delta_u < \infty.
    \label{eq:constraint}
\end{align}

Thus, model~\eqref{eqapp:pageform} contains $2n+1$ unknown parameters: the index of change $j$ and the two sequences of expected values. 
In~\eqref{eq:constraint}, 
if $x_k$ represents the ratio of daily positive cases in a region, the most natural choice is $\delta_\ell=\delta_u=1$, but it is  convenient to consider the general case having an implied hysteresis.
For example, $\delta_u$ may be specified based on tolerable time to reach hospital capacity, while $\delta_\ell$ may be 
based on the time citizens can endure restrictions before reopening the economy or tolerable level of positive cases.  

One might also consider
\begin{align}
&\mu_0 \leq \delta_\ell, \quad 
\mu_1 > \delta_u, \quad 
0 < \delta_\ell \leq \delta_u < \infty, 
\label{eq:peter}
\end{align}
in place of~\eqref{eq:constraint}. In some sense, this might be more natural, since the mean levels before and after the change are still assumed unknown, but are merely constant. However, formulation~\eqref{eq:peter} does not admit a recursive Page-like procedure whereas MAST that results from~\eqref{eq:constraint} does.

According to the Generalized Likelihood Ratio Test (GLRT) principle~\cite{kaydetection,poorbook}, the decision statistic for problem~\eqref{eqapp:pageform} 
is

\vspace*{-7pt}
{\small 
\begin{align} 
& 
\frac
{\displaystyle{\sup_{1 \le j \le n+1, \, \{ \mu_{0,k} \}_{k=1}^{j-1}, \, \{ \mu_{1,k} \}_{k=j}^{n} }}  \;  
\prod_{k=1}^{j-1}  e^{- \frac{(x_k-\mu_{0,k})^2}{2 \sigma^2}}   \prod_{k=j}^{n} e^{- \frac{(x_k-\mu_{1,k})^2}{2 \sigma^2}}}
{ \displaystyle{\sup_{\{ \mu_{0,k} \}_{k=1}^{n} } } \;
{\prod_{k=1}^{n}   e^{- \frac{(x_k-\mu_{0,k})^2}{2 \sigma^2}} }} \nonumber \\& 
=
\frac
{\displaystyle{\sup_{1 \le j \le n+1}}  \;  
\prod_{k=1}^{j-1}  \sup_{\mu_{0,k} \le \delta_\ell} \, e^{- \frac{(x_k-\mu_{0,k})^2}{2 \sigma^2}}  \;  \prod_{k=j}^{n} \sup_{\mu_{1,k} > \delta_u } e^{- \frac{(x_k-\mu_{1,k})^2}{2 \sigma^2}}}
{\displaystyle{\prod_{k=1}^{n}   \sup_{\mu_{0,k} \le \delta_\ell }} \; e^{- \frac{(x_k-\mu_{0,k})^2}{2 \sigma^2}} }, \nonumber 
\end{align}}where the equality follows by recognizing that each factor of the products involves a single value of $\mu_{0,k}$ or $\mu_{1,k}$ and making explicit the constraints in~\eqref{eq:constraint}. The suprema over $\mu_{0,k}$ and $\mu_{1,k}$ appearing in the above expression can be computed in closed form, as follows:
\begin{align}
\sup_{\mu_{0,k} \le \delta_\ell} e^{- \frac{(x_k-\mu_{0,k})^2}{2 \sigma^2}} =
\begin{cases}
e^{- \frac{(x_k-\delta_\ell)^2}{2 \sigma^2}}, & \textnormal{if } x_k > \delta_\ell, \\
1,  & \textnormal{if } x_k \le \delta_\ell,
\end{cases} \label{eqapp:cond1} \\
\sup_{\mu_{1,k}>\delta_u} e^{- \frac{(x_k-\mu_{1,k})^2}{2 \sigma^2}} = 
\begin{cases}
e^{- \frac{(x_k-\delta_u)^2}{2 \sigma^2}}, & \textnormal{if } x_k \le \delta_u, \\
1,  & \textnormal{if } x_k > \delta_u,
\end{cases} \label{eqapp:cond2}
\end{align}which means that the ML (maximum likelihood) estimates of the unknown parameters are, respectively, 
\begin{align}
\widehat \mu_{0,k}=\min(x_k,\delta_\ell), \qquad \widehat \mu_{1,k}=\max(x_k,\delta_u).    
\end{align}
This yields the GLRT statistic in the form
{\small
\begin{align}
&\max_{1\le j \le n+1}
\frac
{\displaystyle{\prod_{1\le k \le j-1 \; : \; \, x_k > \delta_\ell} } e^{- \frac{(x_k-\delta_\ell)^2}{2 \sigma^2}}
\displaystyle{\prod_{j\le k \le n \; : \; \, x_k\le \delta_u} } e^{- \frac{(x_k-\delta_u)^2}{2 \sigma^2}}}
{\displaystyle{\prod_{1\le k \le n \; : \; \, x_k > \delta_\ell} } e^{- \frac{(x_k-\delta_\ell)^2}{2 \sigma^2}}} \nonumber \\
&=\max_{1\le j \le n+1}
\frac
{\displaystyle{\prod_{j\le k \le n \; : \; \, x_k\le \delta_u} } e^{- \frac{(x_k-\delta_u)^2}{2 \sigma^2}}}
{\displaystyle{\prod_{j\le k \le n \; : \; \, x_k > \delta_\ell} } e^{- \frac{(x_k-\delta_\ell)^2}{2 \sigma^2}}} ,
\label{eqapp:page2}
\end{align}}
or, equivalently, taking the logarithm:
\begin{align}
T_n(\delta_\ell,\delta_u)=\max_{1\le j \le n+1}
\overline T_{j:n}(\delta_\ell,\delta_u),
\label{eq:defT}
\end{align}
where 
{\small \begin{align}
\overline T_{j:n}(\delta_\ell,\delta_u)=
\displaystyle{ \sum_{\begin{smallmatrix} k=j \\ k : \, x_k > \delta_\ell \end{smallmatrix}}^n } \frac{(x_k-\delta_\ell)^2}{2 \sigma^2} 
- \hspace*{-10pt}\displaystyle{\sum_{\begin{smallmatrix} k=j \\ k : \, x_k\le \delta_u \end{smallmatrix}}^n } \frac{(x_k-\delta_u)^2}{2 \sigma^2} .
\label{eqapp:page3}
\end{align}}The passage from the controlled to the critical regime is declared at the smallest $n$ such that
\begin{align}
    T_n(\delta_\ell,\delta_u) > \gamma,
    \label{eq:test}
\end{align}
where the threshold level $\gamma$ is selected to trade-off decision delay and risk, two quantities that will be defined in Sec.~\ref{sec:pa}.

The test in~\eqref{eq:test} 
will be referred to as MAST$(\delta_\ell,\delta_u)$
with boundaries $\delta_\ell$ and $\delta_u$. The subscript $n$ appended to $T_n(\delta_\ell,\delta_u)$ denotes its dependence on the stream of data $x_1,\dots,x_n$, and the subscript $j\!:\!n$ appended to $\overline  T_{j:n}(\delta_\ell,\delta_u)$ denotes its dependence on $x_j,\dots,x_n$.
Finally, by introducing the non-linearity
\begin{align}
    g(x_k; \delta_\ell,\delta_u)= \begin{cases}
    -\frac{(x_k-\delta_u)^2}{2 \sigma^2}, & x_k \le \delta_\ell, \\
    \frac{\delta_u-\delta_\ell}{\sigma^2}\left ( x_k - \frac{\delta_\ell+\delta_u}{2}\right ),& \delta_\ell < x_k \le \delta_u, \\
    \frac{(x_k-\delta_\ell)^2}{2 \sigma^2}, & x_k > \delta_u,
    \end{cases}
    \label{eq:gfun}
\end{align}
we have $\overline T_{j:n}(\delta_\ell,\delta_u)=\sum_{k=j}^n g(x_k; \delta_\ell,\delta_u)$.

As a sanity check, let us assume that values of  $x_k$ closer to $\delta_\ell$ are confused with~$\delta_\ell$ and, likewise, values of  $x_k$ closer to $\delta_u$ are confused with~$\delta_u$.
Then, we see from~\eqref{eqapp:page3} that the contribution to $\overline T_{j:n}(\delta_\ell,\delta_u)$ provided by the sample $x_k$ is $\pm(\delta_u-\delta_\ell)^2/2\sigma^2$, where the negative sign applies to the former case and the positive one to the latter.
In the actual operation of $\overline T_{j:n}(\delta_\ell,\delta_u)$, the contribution given by the sample $x_k$ is regulated by its distance to the boundaries, as shown in~\eqref{eq:gfun}: 
\begin{itemize}
    \item values $x_k \le \delta_\ell$ give a negative contribution proportional to the square of the distance of $x_k$ from the upper boundary $\delta_u$;
    \item values $\delta_\ell \le x_k < \delta_u$ give a linear contribution, 
    whose sign depends on which boundary $x_k$ is closest to;
    \item values $x_k > \delta_u$ give a positive contribution proportional to the square of the distance of $x_k$ from the lower boundary $\delta_\ell$.
\end{itemize}

Using the non-linearity of~\eqref{eq:gfun} in~\eqref{eq:defT}, one gets
\begin{align}
T_n(\delta_\ell,\delta_u) &= \max_{1\le j \le n+1} \sum_{k=j}^n g(x_k; \delta_\ell,\delta_u) \nonumber \\
&= \max \left [0, \max_{1\le j \le n} \sum_{k=j}^n g(x_k; \delta_\ell,\delta_u)\right ],
\label{eqapp:pagefinal}
\end{align}
where we have used $\sum_{j=n+1}^{n} g(x_k; \delta_\ell,\delta_u) =0$.

The MAST$(\delta_\ell,\delta_u)$ decision statistic~\eqref{eqapp:pagefinal} can be expressed in recursive form. To see this, let us define  
$S_{m}=\max_{1\le j \le m} \; G_j^m$, with $G_j^m=\sum_{k=j}^m g(x_k; \delta_\ell,\delta_u)$, $m=1,\dots,n$. By using the notation $(x)^+=\max[0,x]$, 
we see that~\eqref{eqapp:pagefinal} can be written as $T_n(\delta_\ell,\delta_u)=(S_n)^+$. Then,
{\small\begin{align}
& T_n(\delta_\ell,\delta_u)=(S_n)^+ = \max \left [0, S_n \right ] 
 =\max \left [0, \max \left [G_1^n, \dots, G_n^n \right ] \right ] \nonumber \\
&  = \max \left [0, g(x_n; \delta_\ell,\delta_u) + \max \left [G_1^{n-1}, \dots, G_{n-1}^{n-1}, 0 \right ] \right ] \nonumber \\
& = \max \left [0, g(x_n; \delta_\ell,\delta_u) + \max \left [\max \left [G_1^{n-1}, \dots, G_{n-1}^{n-1} \right ], 0 \right ] \right ] \nonumber \\
& = \max \left [0, g(x_n; \delta_\ell,\delta_u) + \max \left [S_{n-1}, 0 \right ] \right ] \nonumber \\
&  = \left ( g(x_n; \delta_\ell,\delta_u) + \max \left [S_{n-1}, 0 \right ]\right )^+ 
= \left ( g(x_n; \delta_\ell,\delta_u) + \left (S_{n-1}\right )^+\right )^+ \nonumber \\
& =  \left ( g(x_n; \delta_\ell,\delta_u) + T_{n-1}(\delta_\ell,\delta_u)\right )^+.
\label{eqapp:pagefinal2}
\end{align}}We have thus arrived at a recursive expression for the decision statistic: $T_0(\delta_\ell,\delta_u)=0$ and, for $n \ge 1$, $T_n(\delta_\ell,\delta_u)=g(x_n; \delta_\ell,\delta_u) + T_{n-1}(\delta_\ell,\delta_u)$, if $g(x_n; \delta_\ell,\delta_u) + T_{n-1}(\delta_\ell,\delta_u) \ge 0$, and $T_n(\delta_\ell,\delta_u)=0$, otherwise.
Equivalently: $T_0(\delta_\ell,\delta_u)=0$, and, for $n \ge 1$,
\begin{align}
T_{n}(\delta_\ell,\delta_u)=\max \bigg [ 0, T_{n-1}(\delta_\ell,\delta_u) + g(x_n; \delta_\ell,\delta_u) \bigg ].
\label{eq:pagesquare}
\end{align}

We now consider two special cases. 
First, let $\delta_\ell=\delta_u=\delta$, a case referred to as the MAST$(\delta)$ detector, with decision statistic $T_0(\delta)=0$ and, for $n\ge 1$,
\begin{align}
    T_n(\delta)=\max \bigg [ 0, T_{n-1}(\delta) + 
    \frac{(x_n-\delta)^2}{2 \sigma^2} \sign(x_n-\delta) \bigg ].
    \label{eq:MASTdelta}
\end{align}
Further assuming $\delta=1$ in~\eqref{eq:MASTdelta}, yields a decision procedure that we simply call MAST, 
whose decision statistic $T_n(1)$ is denoted by $T_n$: $T_0=0$ and, for $n\ge 1$,
\begin{align}
    T_n=\max \bigg [ 0, T_{n-1} + 
    \frac{(x_n-1)^2}{2 \sigma^2} \sign(x_k-1) \bigg ].
    \label{eq:MAST}
\end{align}

The second special case is when $\delta_\ell=1-\alpha$ and $\delta_u=1+\alpha$, for some $0<\alpha<1$, which is relevant in connection to Page's test, as discussed next.
As is well-known, if the mean values of the observed sequence before and after the change are \emph{constant and known}, say $\mu_{0,n}=1-\alpha$ and $\mu_{1,n}=1+\alpha$,
the statistic to be compared to a suitable threshold level would be the 
CUSUM~\cite{page,poor-book-quickest,basseville-book}: $Q_0=0$ and, for $n\ge1$,
\begin{align}
Q_{n}=\max \bigg \{ 0, Q_{n-1} + \frac{2 \alpha \, (x_n-1) }{\sigma^2}  \bigg \}. 
\label{eq:page}
\end{align}
For $1-\alpha \le x_k \le 1+\alpha$, Eq.~\eqref{eq:gfun} gives $g(x_k;1-\alpha,1+\alpha)=\frac{2 \alpha}{\sigma^2} (x_k-1)$, which shows that the decision statistic $T_n(1-\alpha,1+\alpha)$ in~\eqref{eqapp:pagefinal} operates exactly as the Page's test for samples $x_k \in [1-\alpha,1+\alpha]$. 

Different optimality criteria have been advocated for the CUSUM test. The ``first-order'' criterion considers the asymptotic situation in which the mean time between false alarms goes to infinity and asserts that  the CUSUM minimizes the worst-case mean delay, where the qualification ``worst'' refers to both the change time and the behavior of the process before change~\cite[p. 166]{basseville-book}.
The test based on~\eqref{eq:page} is in this sense the optimal quickest-detection Page's test.

It is worth noting that the MAST statistic in~\eqref{eq:MAST} is formally obtained by replacing the unknown value of $\alpha$ appearing in the CUSUM statistic, with an estimate $\widehat \alpha_n=|x_n-1|$ (constant factors can be incorporated in the threshold). This suggests an analogy between MAST for quickest-detection problems and the energy detector for testing the presence of an unknown time-varying deterministic signal buried in Gaussian noise, in the classical hypothesis testing framework~\cite{kaydetection}.

\section{Performance Assessment} 
\label{sec:pa}
The performance of MAST$(\delta_\ell,\delta_u)$ is expressed in terms of mean delay time $\Delta$ and the risk $R$. The mean delay $\Delta$ is the difference between the time at which the MAST$(\delta_\ell,\delta_u)$ statistic $T_n(\delta_\ell,\delta_u)$ crosses a preassigned threshold level $\gamma$, see~\eqref{eq:test}, and the time of passage from the controlled to the critical regime. 
In the critical regime, the pandemic grows exponentially fast and it is therefore important to ensure that $\Delta$ be as small as possible.
This requirement is in contrast with the requirement $R \ll 1$. The risk $R$ is defined as the reciprocal of the mean time between two false alarms\footnote{Note that in a 
quickest detection application the concept of a ``false alarm'' is different from that in a fixed-block test.}. In turn, the mean time between false alarms is the mean time between two threshold crossings, assuming that the decision statistic is reset to zero at any threshold crossing event, occurring in the controlled regime. Because of the unwelcome social and economic impact of the measures presumably taken by the authorities when passage into the critical regime is detected, it is evident that~$R$ must be extremely small. The same performance indices~$\Delta$ and~$R$ used to characterize MAST$(\delta_\ell,\delta_u)$ are used for the Page's test.

We now investigate the performance of MAST$(\delta_\ell,\delta_u)$ by computer experiments, limiting the analysis to the case $\delta_\ell=\delta_u=1$, i.e., the simple MAST. The performance of the Page's test is used as a benchmark. 
Let us consider the following ``scenario~0''. Fix $\alpha>0$. Suppose that the state of nature (mean value of the $x_n$'s) is $\mu_{0,n}=1-\alpha$ for all $n$ in the controlled regime; likewise, suppose $\mu_{1,n}=1+\alpha$ for all $n$ in the critical regime. By standard Monte Carlo counting, for MAST we found that the delay $\Delta$ varies almost linearly with the threshold level $\gamma$, and that $\log_{10} R$ varies almost linearly with~$\gamma$. 
The same approximate behavior is found, again by standard Monte Carlo counting, for the \emph{clairvoyant} Page's test that is aware of the mean values $\mu_{0,n}=1-\alpha$ and $\mu_{1,n}=1+\alpha$: the mappings $\gamma \mapsto \Delta$ and $\gamma \mapsto \log_{10} R$ are approximately linear. These numerical analyses are not detailed for the sake of brevity. 
The observed behavior is known for the Page's test, at least when the threshold $\gamma$ is sufficiently large, in view of the Wald's approximation, see, e.g.~\cite[Eq.\ 5.2.44]{basseville-book}. 
In the present Gaussian case, more accurate formulas --- known as Siegmund's approximations --- are also available~\cite[Eqs.\ 5.2.64, 5.2.65]{basseville-book}.

\begin{figure}
\centering 
\hspace{10mm}\psfragfig[width=.90\columnwidth,height=.6\columnwidth]{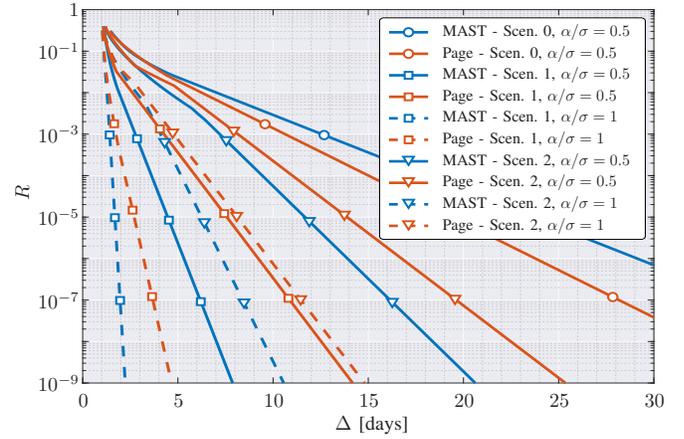}
\vspace{2mm}
\caption{Operational characteristic (risk $R$ versus decision delay $\Delta$) of the MAST quickest detection test, compared to the benchmark Page's test. Three scenarios are considered, as described in the main text.
In scenario~0, Page's test is optimal. 
MAST outperforms Page's test in scenarios~1 and 2, in which the sequences $\{\mu_{0,n}\}$ and $\{\mu_{1,n}\}$ are time-varying. Scenario~2, in particular, mimics the actual behavior of the sequences, as observed in COVID-19 pandemic data, see Sec.~\ref{sec:COVID}.}
\label{fig:tmp}
\end{figure}

We assume that the aforementioned linear mappings observed for MAST and Page's test hold true for any value of the threshold, 
and this assumption allows us to consider values of the mean delay and (especially) of the risk that would be difficult to obtain by standard Monte Carlo analysis. 
In this way, we obtain the operational curve of the two decision systems shown in Fig.~\ref{fig:tmp}. The operational curve is the relationship between
$R$ and~$\Delta$. As expected, Page's test outperforms the MAST, because the Page's test is optimal for the case addressed in scenario~0. 

The same numerical analysis has been conducted for ``scenario 1'' and ``scenario 2'', also shown in Fig.~\ref{fig:tmp}.
In scenario~1, we suppose that in the controlled regime, any $\mu_{0,n}$ is an instantiation of a uniform random variable with support $(1-\alpha,1)$, while in the critical regime any $\mu_{1,n}$ is an instantiation of a uniform random variable with support $(1,1+10 \, \alpha)$.
In scenario~2, instead, we suppose that the sequences $\{ \mu_{0,k} \}$ and $\{ \mu_{1,k} \}$ are sinusoidal with a period of 75 days.\footnote{
Scenario~2 is consistent with the sequences of mean values obtained by the COVID-19 epidemic data observed for different countries~\cite{MAST-Nature}.}
Specifically, in the controlled regime the sinusoid oscillates in $(1 - \alpha,1)$, while in the critical regime it oscillates in $(1,1 + 10 \, \alpha)$.
To implement the Page's test in both scenarios~1 and~2, it is assumed that the mean values are constant, i.e., $\mu_{0,n}=1-\alpha$ and $\mu_{1,n}=1+\alpha$, as in scenario~0.
Clearly, no assumption about the mean values is instead needed for implementing the MAST test, except that they are bounded by one. In Fig.~\ref{fig:tmp}, we see that MAST outperforms Page's test, confirming its effectiveness when the mean values $\{\mu_{0,n}\}$ and $\{\mu_{1,n}\}$ are unknown, except for being bounded as shown in~\eqref{eq:constraint}.

\section{Application to COVID-19 pandemic data}
\label{sec:COVID}

Starting from the landmark SIR model developed in~\cite{KerMckWal:J27}, a multitude of sophisticated epidemiological models have been proposed to describe the pandemic evolution, based, e.g., on stochastic evolution of epidemic compartments~\cite{SkvRis:J12,HuQiaJunXiaWeiZhi:J20,MaiBenBro:K20,AdaptiveBayesian,Allen2017}, or metapopulation networks,~\cite{Li489, Chinazzi395}, just to cite two examples. The trend in the topical literature is to conceive increasingly complex models, often suitable for analysis by big-data techniques. The main goal of these models is to predict mid/long-term evolution of the infection. 
Our focus, instead, is to quickly detect the onset of the exponential growth. With this aim, we consider an abbreviated observation model, built on the concept that the pandemic evolution is essentially a multiplicative phenomenon. 

We model the number of new positive individuals on day~$n$, 
say~$p_n$, as the number $p_{n-1}$ of new positive individuals on day~$n-1$, multiplied by a random variable~$x_n$. Further including a ``noise'' term $w_n$, yields the scalar discrete-time state equation $p_n= p_{n-1} x_n + w_n$,  $n\ge 1$,
for some initial state~$p_0$. Such a recursion,
under various assumptions for the sequences $\{(x_n, w_n)\}$, 
is known as a \emph{perpetuity} and appears in many disciplines~\cite{vervaat1979,EmbrechtsGoldie94,renormingperpetuities}. 
We assume that the noise term $w_n$
is negligible, 
yielding:\footnote{The same multiplicative structure shown in~\eqref{eq:multiplicative} applies, other than $p_n$, to different time-series related to the pandemic evolution, e.g., the number of new hospitalizations per day~\cite{MAST-Nature}.}  
\begin{align}
p_n=p_{n-1} x_n \quad \Rightarrow \quad
    p_n=p_0 \prod_{k=1}^n x_k,
    \label{eq:multiplicative}
\end{align} 
for some $p_0>0$.
In this article, we refer to model~\eqref{eq:multiplicative}, in which $x_1,x_2,\dots$ are independent random variables. This is akin to the popular random walk model, with the independence of the increments of the random walk replaced by the independence of the ratios $p_{n}/p_{n-1}$.
Model~\eqref{eq:multiplicative} is derived from SIR-like models and validated on COVID-19 data in~\cite{MAST-Nature}, where it is also 
shown that the $x_n$'s closely follow a Gaussian distribution with (unknown) time-varying expected value $\E x_n$, and a common standard deviation\footnote{Since $\sigma \ll 1$ and $\E x_n \approx 1$, $\P(x_n<0)$ is negligible, for all $n$. Thus, one can safely assume that $\{x_n\}$ is a sequence of independent \emph{nonnegative} random variables.}~$\sigma$.

As long as $\E x_n<1$, the sequence $\{p_n\}$ tends to decay exponentially to zero, while, for $\E x_n >1$, $\{p_n\}$ tends to increase exponentially fast. We are interested in quickly detecting the passage from the former situation (a controlled regime) to the latter (critical). Detecting this change can be cast in terms of a binary decision problem between two hypotheses, referred to as the \emph{null} and the \emph{alternative}, as 
shown in~\eqref{eqapp:pageform}.

\begin{figure}
\centering 
\hspace{10mm}\psfragfig[width=.90\columnwidth,height=.6\columnwidth]{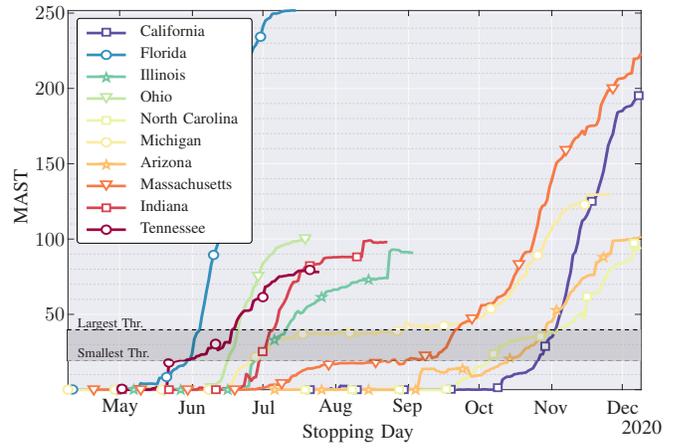}
\vspace{2.8mm}
\caption{MAST decision statistic computed for 10 US states and used to detect the onset of the COVID-19 second wave.
The dashed horizontal lines represent the smallest and largest thresholds corresponding to $R = 10^{-9}$, for the ensemble of the ten states. Curves are prolonged beyond threshold crossing for clarity.}
\label{fig:MAST}
\end{figure}

An example of application of MAST to COVID-19 data is provided in Fig.~\ref{fig:MAST}. The abscissa point at which the MAST statistic crosses the threshold represents the day at which the onset is detected.
The test threshold is state-dependent, as discussed in~\cite{MAST-Nature}.
Then, for clarity of illustration, only the smallest and largest thresholds corresponding to the risk $R = 10^{-9}$ are shown, which for many states makes only a few days difference as to the time of alert.
One observation is that restrictive measures have not been adopted in as timely a manner as suggested by the MAST analysis.
The reader is referred to~\cite{MAST-Nature,millefiori2020covid19,soldi2021commag,MarSay-submitted,WEBSITE} for details.
Several aspects of the MAST analysis of COVID-19 data deserve further study. These include the pre-processing to clean the data from gross errors (e.g., asynchronous or unreported data); generalization of the approach to analyze other publicly available time-series (e.g., number of hospitalized, number of deaths)
and even as a vector of observations; on-line estimation of the variance to make the detector robust to statistical fluctuations, often observed in COVID-19 data.

\section{Conclusion}

This article derived a sequential test called MAST, which is used in~\cite{MAST-Nature} to detect passage from the controlled regime in which the COVID-19 pandemic is restrained, to the critical regime in which the infection spreads exponentially fast. MAST is a variation of the celebrated Page's test based on the CUSUM statistic, designed for cases in which the expected values of the data are bounded below a lower barrier~$\delta_\ell$ in the controlled regime, and above an upper
barrier~$\delta_u$ in the critical one, but are otherwise unknown.
We show that MAST admits a recursive form and in the simplest case $\delta_\ell=\delta_u=1$, is formally obtained from the Page's test with nominal expected values $1 \pm \alpha$, by replacing $\alpha$ with an estimate thereof. 
The performance of MAST is investigated by computer experiments. 
If the expected values of the data are constant and known, the performance loss of MAST with respect to the optimal Page's test is moderate. In pandemic scenarios, lacking knowledge of the expected values of the data, MAST can well overcome the Page's test designed with nominal values of the unknowns.

\clearpage
\balance
\bibliographystyle{IEEEtran}
\bibliography{IEEEabrv,mybib_upd_clean}

\begin{thebibliography}{10}
\providecommand{\url}[1]{#1}
\csname url@samestyle\endcsname
\providecommand{\newblock}{\relax}
\providecommand{\bibinfo}[2]{#2}
\providecommand{\BIBentrySTDinterwordspacing}{\spaceskip=0pt\relax}
\providecommand{\BIBentryALTinterwordstretchfactor}{4}
\providecommand{\BIBentryALTinterwordspacing}{\spaceskip=\fontdimen2\font plus
\BIBentryALTinterwordstretchfactor\fontdimen3\font minus
  \fontdimen4\font\relax}
\providecommand{\BIBforeignlanguage}[2]{{%
\expandafter\ifx\csname l@#1\endcsname\relax
\typeout{** WARNING: IEEEtran.bst: No hyphenation pattern has been}%
\typeout{** loaded for the language `#1'. Using the pattern for}%
\typeout{** the default language instead.}%
\else
\language=\csname l@#1\endcsname
\fi
#2}}
\providecommand{\BIBdecl}{\relax}
\BIBdecl

\bibitem{page}
E.~Page, ``Continuous inspection schemes,'' \emph{Biometrika}, vol.~41, pp.
  100--115, Jan. 1954.

\bibitem{basseville-book}
M.~Basseville and I.~V. Nikiforov, \emph{Detection of abrupt changes: theory
  and application}.\hskip 1em plus 0.5em minus 0.4em\relax Prentice Hall
  Englewood Cliffs, 1993, vol. 104.

\bibitem{poor-book-quickest}
H.~V. Poor and O.~Hadjiliadis, \emph{Quickest Detection}.\hskip 1em plus 0.5em
  minus 0.4em\relax Cambridge, UK: Cambridge University Press, 2009.

\bibitem{TRUONG2020107299}
C.~Truong, L.~Oudre, and N.~Vayatis, ``Selective review of offline change point
  detection methods,'' \emph{Signal Processing}, vol. 167, p. 107299, 2020.

\bibitem{roberts2021common}
M.~Roberts, D.~Driggs, M.~Thorpe, J.~Gilbey, M.~Yeung, S.~Ursprung, A.~I.
  Aviles-Rivero, C.~Etmann, C.~McCague, L.~Beer, J.~R. Weir-McCall, Z.~Teng,
  E.~Gkrania-Klotsas, J.~H.~F. Rudd, E.~Sala, and C.-B. Schönlieb, ``Common
  pitfalls and recommendations for using machine learning to detect and
  prognosticate for {COVID}-19 using chest radiographs and {CT} scans,'' 2021.

\bibitem{hu2020access}
S.~{Hu}, Y.~{Gao}, Z.~{Niu}, Y.~{Jiang}, L.~{Li}, X.~{Xiao}, M.~{Wang}, E.~F.
  {Fang}, W.~{Menpes-Smith}, J.~{Xia}, H.~{Ye}, and G.~{Yang}, ``Weakly
  supervised deep learning for covid-19 infection detection and classification
  from ct images,'' \emph{IEEE Access}, vol.~8, pp. 118\,869--118\,883, 2020.

\bibitem{Anderson2020}
R.~M. Anderson, H.~Heesterbeek, D.~Klinkenberg, and T.~D. Hollingsworth,
  ``\BIBforeignlanguage{en}{How will country-based mitigation measures
  influence the course of the {COVID}-19 epidemic?}''
  \emph{\BIBforeignlanguage{en}{The Lancet}}, vol. 395, no. 10228, pp.
  931--934, Mar. 2020.

\bibitem{Hellewell2020}
J.~Hellewell, S.~Abbott, A.~Gimma, N.~I. Bosse, C.~I. Jarvis, T.~W. Russell,
  J.~D. Munday, A.~J. Kucharski, W.~J. Edmunds, S.~Funk, R.~M. Eggo, F.~Sun,
  S.~Flasche, B.~J. Quilty, N.~Davies, Y.~Liu, S.~Clifford, P.~Klepac, M.~Jit,
  C.~Diamond, H.~Gibbs, and K.~van Zandvoort, ``Feasibility of controlling
  {COVID}-19 outbreaks by isolation of cases and contacts,'' \emph{The Lancet
  Global Health}, vol.~8, no.~4, pp. e488--e496, Apr. 2020.

\bibitem{Maria2020}
M.~Nicola, Z.~Alsafi, C.~Sohrabi, A.~Kerwan, A.~Al-Jabir, C.~Iosifidis,
  M.~Agha, and R.~Agha, ``\BIBforeignlanguage{en}{The socio-economic
  implications of the coronavirus pandemic ({COVID}-19): {A} review},''
  \emph{\BIBforeignlanguage{en}{International Journal of Surgery}}, vol.~78,
  pp. 185--193, Jun. 2020.

\bibitem{SHARIF2020}
A.~Sharif, C.~Aloui, and L.~Yarovaya, ``{COVID}-19 pandemic, oil prices, stock
  market, geopolitical risk and policy uncertainty nexus in the {US} economy:
  {F}resh evidence from the wavelet-based approach,'' \emph{International
  Review of Financial Analysis}, vol.~70, p. 101496, 2020.

\bibitem{GlobalSupplyChain}
D.~Guan, D.~Wang, S.~Hallegatte, S.~J. Davis, J.~Huo, S.~Li, Y.~Bai, T.~Lei,
  Q.~Xue, D.~Coffman, D.~Cheng, P.~Chen, X.~Liang, B.~Xu, X.~Lu, S.~Wang,
  K.~Hubacek, and P.~Gong, ``Global supply-chain effects of {COVID}-19 control
  measures,'' \emph{Nature Human Behaviour}, vol.~4, no.~6, pp. 577--587, Jun.
  2020.

\bibitem{millefiori2020covid19}
L.~M. Millefiori, P.~Braca, D.~Zissis, G.~Spiliopoulos, S.~Marano, P.~K.
  Willett, and S.~Carniel, ``{COVID}-19 impact on global maritime mobility,''
  2020.

\bibitem{tartakovsky-book}
A.~Tartakovsky, I.~Nikiforov, and M.~Basseville, \emph{Sequential analysis:
  Hypothesis testing and changepoint detection}.\hskip 1em plus 0.5em minus
  0.4em\relax CRC Press, 2014.

\bibitem{kaydetection}
S.~M. Kay, \emph{Fundamentals of Statistical Signal Processing, Volume II:
  Detection Theory}.\hskip 1em plus 0.5em minus 0.4em\relax Prentice Hall PTR,
  1998.

\bibitem{poorbook}
H.~V. Poor, \emph{An Introduction to Signal Detection and Estimation}.\hskip
  1em plus 0.5em minus 0.4em\relax New York: Springer-Verlag, 1988.

\bibitem{MAST-Nature}
P.~Braca, D.~Gaglione, S.~Marano, L.~M. Millefiori, P.~Willett, and
  K.~Pattipati, ``Quickest detection of {Critical} covid-19 phases: When should
  restrictive measures be taken?'' 2020.

\bibitem{KerMckWal:J27}
W.~O. Kermack, A.~G. McKendrick, and G.~T. Walker, ``A contribution to the
  mathematical theory of epidemics,'' \emph{Proc. R. Soc. Lond}, vol. 115, no.
  772, pp. 700--721, Aug. 1927.

\bibitem{SkvRis:J12}
A.~Skvortsov and B.~Ristic, ``Monitoring and prediction of an epidemic outbreak
  using syndromic observations,'' \emph{Math. Biosci.}, vol. 240, no.~1, pp.
  12--19, Nov. 2012.

\bibitem{HuQiaJunXiaWeiZhi:J20}
Z.~Hu, Q.~Cui, J.~Han, X.~Wang, W.~E. Sha, and Z.~Teng, ``Evaluation and
  prediction of the {COVID-19} variations at different input population and
  quarantine strategies, a case study in {Guangdong} province, {China},''
  \emph{Int. J. Infect. Dis.}, vol.~95, pp. 231--240, Jun. 2020.

\bibitem{MaiBenBro:K20}
B.~F. Maier and D.~Brockmann, ``Effective containment explains subexponential
  growth in recent confirmed {COVID-19} cases in {China},'' \emph{Science},
  vol. 368, no. 6492, pp. 742--746, May 2020.

\bibitem{AdaptiveBayesian}
D.~Gaglione, P.~Braca, L.~M. Millefiori, G.~Soldi, N.~Forti, S.~Marano,
  P.~Willett, and K.~R. Pattipati, ``Adaptive {B}ayesian learning and
  forecasting of epidemic evolution - {D}ata analysis of the {COVID}-19
  outbreak,'' \emph{IEEE Access}, vol.~8, pp. 175\,244--175\,264, 2020.

\bibitem{Allen2017}
L.~J. Allen, ``A primer on stochastic epidemic models: Formulation, numerical
  simulation, and analysis,'' \emph{Infectious Disease Modelling}, vol.~2,
  no.~2, pp. 128--142, may 2017.

\bibitem{Li489}
R.~Li, S.~Pei, B.~Chen, Y.~Song, T.~Zhang, W.~Yang, and J.~Shaman,
  ``Substantial undocumented infection facilitates the rapid dissemination of
  novel {Coronavirus} {(SARS-CoV-2)},'' \emph{Science}, vol. 368, no. 6490, pp.
  489--493, May 2020.

\bibitem{Chinazzi395}
M.~Chinazzi, J.~T. Davis, M.~Ajelli, C.~Gioannini, M.~Litvinova, S.~Merler,
  A.~Pastore~y Piontti, K.~Mu, L.~Rossi, K.~Sun, C.~Viboud, X.~Xiong, H.~Yu,
  M.~E. Halloran, I.~M. Longini, and A.~Vespignani, ``The effect of travel
  restrictions on the spread of the 2019 novel {Coronavirus} {(COVID-19)}
  outbreak,'' \emph{Science}, vol. 368, no. 6489, pp. 395--400, Apr. 2020.

\bibitem{vervaat1979}
W.~Vervaat, ``On a stochastic difference equation and a representation of
  non-negative infinitely divisible random variables,'' \emph{Adv. Appl. Prob},
  vol.~11, pp. 750--783, 1979.

\bibitem{EmbrechtsGoldie94}
P.~Embrechts and C.~Goldie, ``Perpetuities and random equations,'' in
  \emph{Asymptotic Statistics}, P.~M. {et al.}, Ed.\hskip 1em plus 0.5em minus
  0.4em\relax Berlin Heidelberg: Springer-Verlag, 1994, pp. 75--86.

\bibitem{renormingperpetuities}
P.~Hitczenko and J.~Wesolowski, ``Renorming divergent perpetuities,''
  \emph{Bernoulli}, vol.~17, no.~3, pp. 880--894, 2011.

\bibitem{soldi2021commag}
G.~Soldi, N.~Forti, D.~Gaglione, P.~Braca, L.~M. Millefiori, S.~Marano,
  P.~Willett, and K.~Pattipati, ``Quickest detection and forecast of pandemic
  outbreaks: Analysis of {COVID}-19 waves,'' 2021.

\bibitem{MarSay-submitted}
S.~Marano and A.~H. Sayed, ``Decision-making algorithms for learning and
  adaptation with application to {COVID-19} data,'' 2020.

\bibitem{WEBSITE}
P.~Braca, D.~Gaglione, S.~Marano, L.~M. Millefiori, P.~Willett, and
  K.~Pattipati, ``{MAST}: {COVID-19} pandemic onset test -- {M}ulti-country
  analysis and visualization,'' \url{https://covid-mast.github.io}, 2020.

\end{thebibliography}

\end{document}